%% file: icassp2020.tex
\documentclass{article}
\usepackage{spconf,amsmath,graphicx}
\usepackage{soul}
\usepackage{url}
\usepackage[utf8]{inputenc}
\usepackage{graphicx}
\usepackage{amsmath}
\usepackage{booktabs}
\usepackage{algorithm}
\usepackage{algorithmic}
\usepackage{mathtools}
\usepackage{amsfonts}
\usepackage{blindtext}
\usepackage{booktabs}
\usepackage{multirow}

\title{MULTIMODAL LEARNING FOR CLASSROOM ACTIVITY DETECTION}
\name{Hang Li, Yu Kang, Wenbiao Ding, Song Yang, Songfan Yang, Gale Yan Huang, Zitao Liu\sthanks{The corresponding author. Email: liuzitao@100tal.com.}}
\address{TAL AI Lab, TAL Education Group, Beijing, China}
\begin{document}
%
\maketitle
\begin{abstract}
\input{abstract}

\end{abstract}


\section{Introduction}
\input{intro}

\section{Related Work}
\input{related}

\section{Our Approach}
\input{method}

\section{Experiments}
\input{experiment}

\section{Conclusion}
\input{conclusion}

\vfill\pagebreak

\small
\bibliographystyle{IEEEbib}
\bibliography{icassp2020}

\end{document}

%% file: abstract.tex
Classroom activity detection (CAD) focuses on accurately classifying whether the teacher or student is speaking and recording both the length of individual utterances during a class. A CAD solution helps teachers get instant feedback on their pedagogical instructions. This greatly improves educators' teaching skills and hence leads to students' achievement. However, CAD is very challenging because (1) the CAD model needs to be generalized well enough for different teachers and students; (2) data from both vocal and language modalities has to be wisely fused so that they can be complementary; and (3) the solution shouldn't heavily rely on additional recording device. In this paper, we address the above challenges by using a novel attention based neural framework. Our framework not only extracts both speech and language information, but utilizes attention mechanism to capture long-term semantic dependence. Our framework is device-free and is able to take any classroom recording as input. The proposed CAD learning framework is evaluated in two real-world education applications. The experimental results demonstrate the benefits of our approach on learning attention based neural network from classroom data with different modalities, and show our approach is able to outperform state-of-the-art baselines in terms of various evaluation metrics.

%% file: intro.tex
Teacher-student interaction analysis in live classrooms with the goal of accurately quantifying classroom activities, such as lecturing, discussion, etc is very crucial for student achievement \cite{nystrand2006research,owens2017classroom,d2015multimodal,wang2014automatic,ganek2016language,chen2019multimodal,ding2020dolphin}. It not only provides students the opportunity to work through their understanding and learn from ideas of others but gives teachers epistemic feedback on their instruction which is important for crafting their teaching skills \cite{donnelly2016automatic,donnelly2016multi,olney2017assessing,liu2019automatic}. Such analysis usually takes into account a classroom recording (e.g., as collected by an audio or video recorder) and outputs pedagogical annotations of classroom activities. 

The majority of the current practices of classroom dialogic analysis is logistically complex and expensive, requiring observer rubrics, observer training, and continuous assessment to maintain a pool of qualified observers \cite{archer2016better,olney2017assessing}. Even with performance support tools developed by Nystrand and colleagues, including live coding CLASS 4.25 software \cite{nystrand2006research,nystrand}, it still requires approximately 4 hours of coding time per 1 hour of classroom observation. This is an unsustainable task for scalable research, let alone for providing day-to-day feedback for teacher professional development.

\begin{figure}[!tpbh]
\centering
\includegraphics[width=0.48\textwidth] {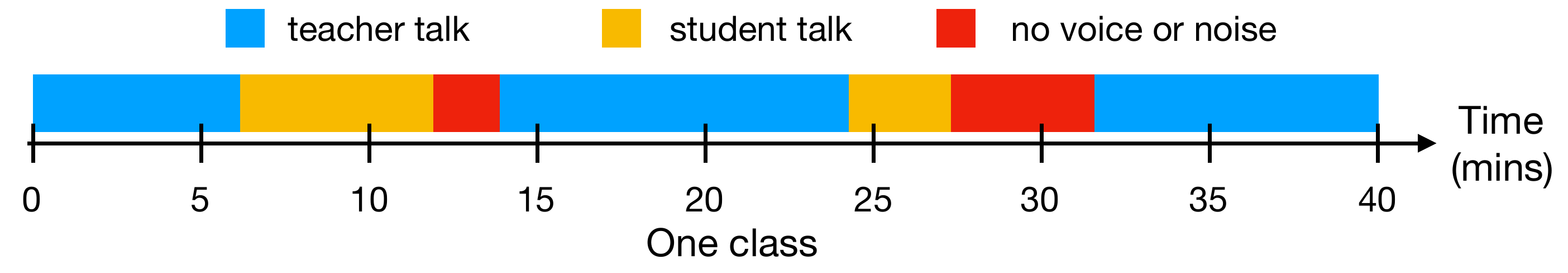} 
\caption{A graphical illustration of CAD results in a sample class. The x-axis represents time within the class.}
\label{fig:example}
\end{figure}

In this work, we focus on the very fundamental and classic classroom activity detection (CAD) problem and aim to automatically distinguish between whether the teacher or student is speaking and record both the length of individual utterances and the total amount of talk during a class. An example annotation trace for a class is illustrated in Figure \ref{fig:example}. The CAD results of identified activity patterns during lessons will give valuable information about the quantity and distribution of classroom talk and therefore help teachers improve their interactions with students so as to improve student achievement.

A large spectrum of models have been developed and successfully applied in solving CAD problems \cite{mu2012acodea,owens2017classroom,cosbey2019deep,donnelly2016automatic}. However, CAD in real-world scenarios poses numerous challenges. First, vocal information is usually not enough when solving the CAD task due to the multimodal classroom environment. The teacher's voice might be very close to some student's voice, which undoubtedly poses a hard modeling problem since the existing well-developed approaches either focus on identifying each individual speaker in the clean environment or utilize extra recording devices for such activity detection. Second, teacher-student conversations from real-world classroom scenarios are very causal and open-ended. It is difficult to capture the latent semantic information and how to model the long-term sentence-level dependence remains a big concern. Third, the CAD solution should be flexible and doesn't rely on additional recording devices like portable microphones for only collecting teacher audio.

In this paper we study and develop a novel solution to CAD problems that is applicable and can learn neural network models from the real-world multimodal classroom environment. More specifically, we present an attention based multimodal learning framework which (1) fuses the multimodal information by attention based networks such that the teachers' or students' semantic ambiguities can be alleviated by vocal attention scores; and (2) directly learns and predicts from classroom recordings and doesn't rely on any additional recording device for teachers.

%% file: related.tex
There is a long research history on the use of audio (and video) to study instructional practices and student behaviors in live classrooms and many approaches and schemes are designed for CAD annotations due to different purposes \cite{mu2012acodea,owens2017classroom,cosbey2019deep,donnelly2016automatic}. For examples, Owens et al. develop Decibel Analysis for Research in Teaching, i.e., DART, to analyzes the volume and variance of classroom recordings to predict the quantity of time spend on single voice (e.g., lecture), multiple voice (e.g., pair discussion), and no voice (e.g., clicker question thinking) activities \cite{owens2017classroom}. Cosbey et al. improve the DART performance by using deep and recurrent neural network (RNN) architectures \cite{cosbey2019deep}. A comprehensive comparison experiments of deep neural network (DNN), gated recurrent network (GRU) and RNN are studied. Mu et al. present the Automatic Classification of Online Discussions with Extracted Attributes, i.e., ACODEA, framework for fully automatic segmentation and classification of online discussion \cite{mu2012acodea}. ACODEA focuses on learners' argumentation knowledge acquiring and categorizes the content based on the micro-argumentation dimensions such as Claim, Ground, Warrant, Inadequate Claim, etc \cite{weinberger2006framework}. Donnelly et al. aim to provide teachers with formative feedback on their instructions by training Naive Bayes models to identify occurrences of some key instructional segments, such as Question \& Answer, Procedures and Directions, Supervised Seatwork etc \cite{donnelly2016automatic}. 

The closest related work is research by Wang et. al \cite{wang2014automatic}, who conduct CAD by using LENA system \cite{ganek2016language} and identifies three discourse activities of teacher lecturing, class discussion and student group work. Our work is different from Wang et. al since we develop a novel attention based multimodal neural framework to conduct the CAD tasks in the real-world device-free environment. While Wang et. al need to ask teachers to wear the LENA system during the entire teaching process and use differences in volume and pitch in order to assess when teachers were speaking or students were speaking.

Please note that CAD is different from the classic speaker verification \cite{wan2018generalized,rahman2018attention,heigold2016end} and speaker diarization \cite{wang2018speaker} where (1) there is no enrollment-verification 2-stage process in CAD tasks; and (2) not every speaker need to be identified.

%% file: method.tex
\subsection{Problem Statement}
\input{statement}

\subsection{The Proposed Framework}

\subsubsection{Multimodal Attention Layer}

\begin{figure}[!tpbh]
\begin{minipage}{.475\linewidth}
    \centering
    \centerline{\includegraphics[width=4cm,height=4.2cm] {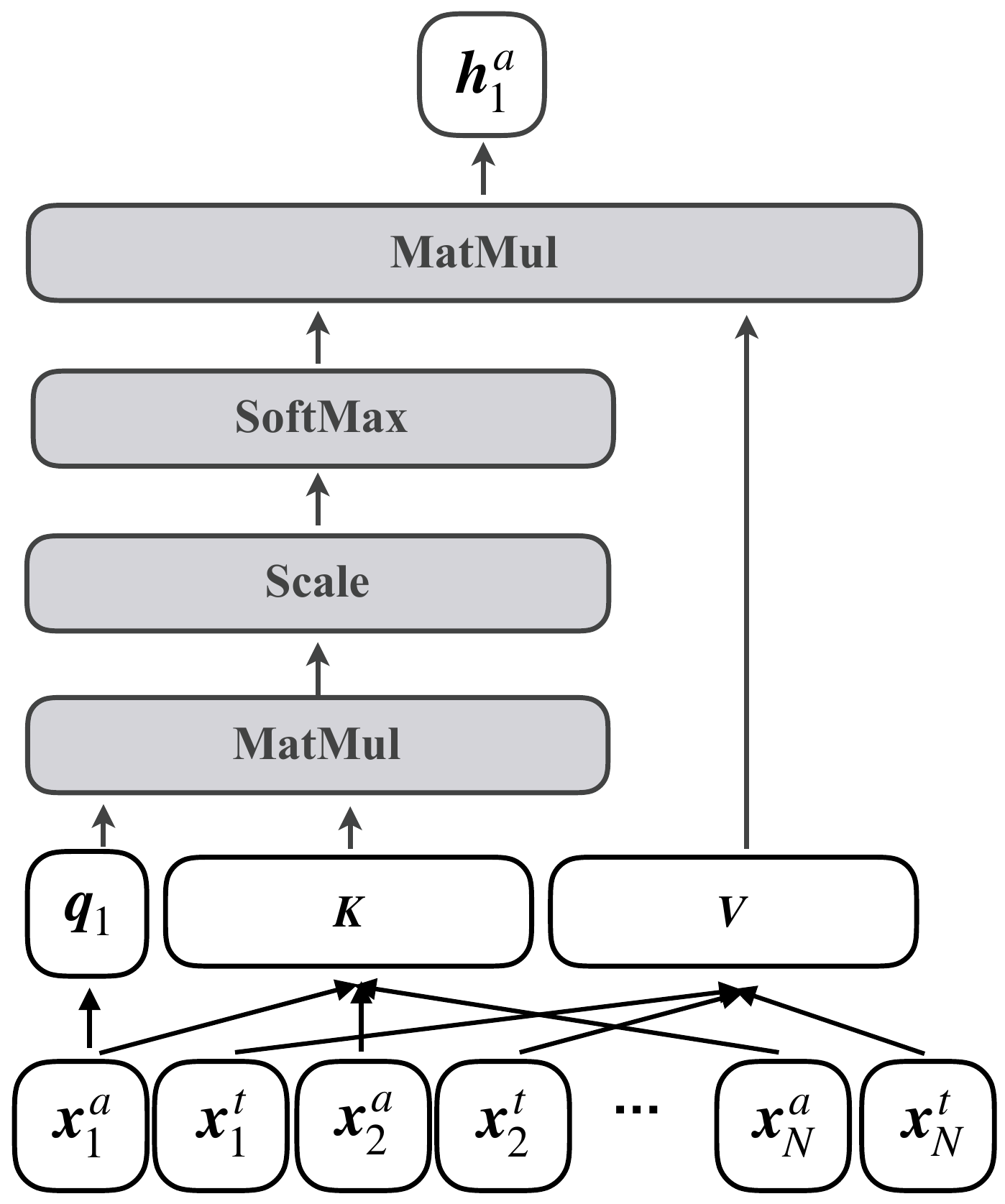}}
    \centerline{(a)}\medskip
\end{minipage}
\hfil
\begin{minipage}{.475\linewidth}
    \centering
    \centerline{\includegraphics[width=4cm,height=4.2cm] {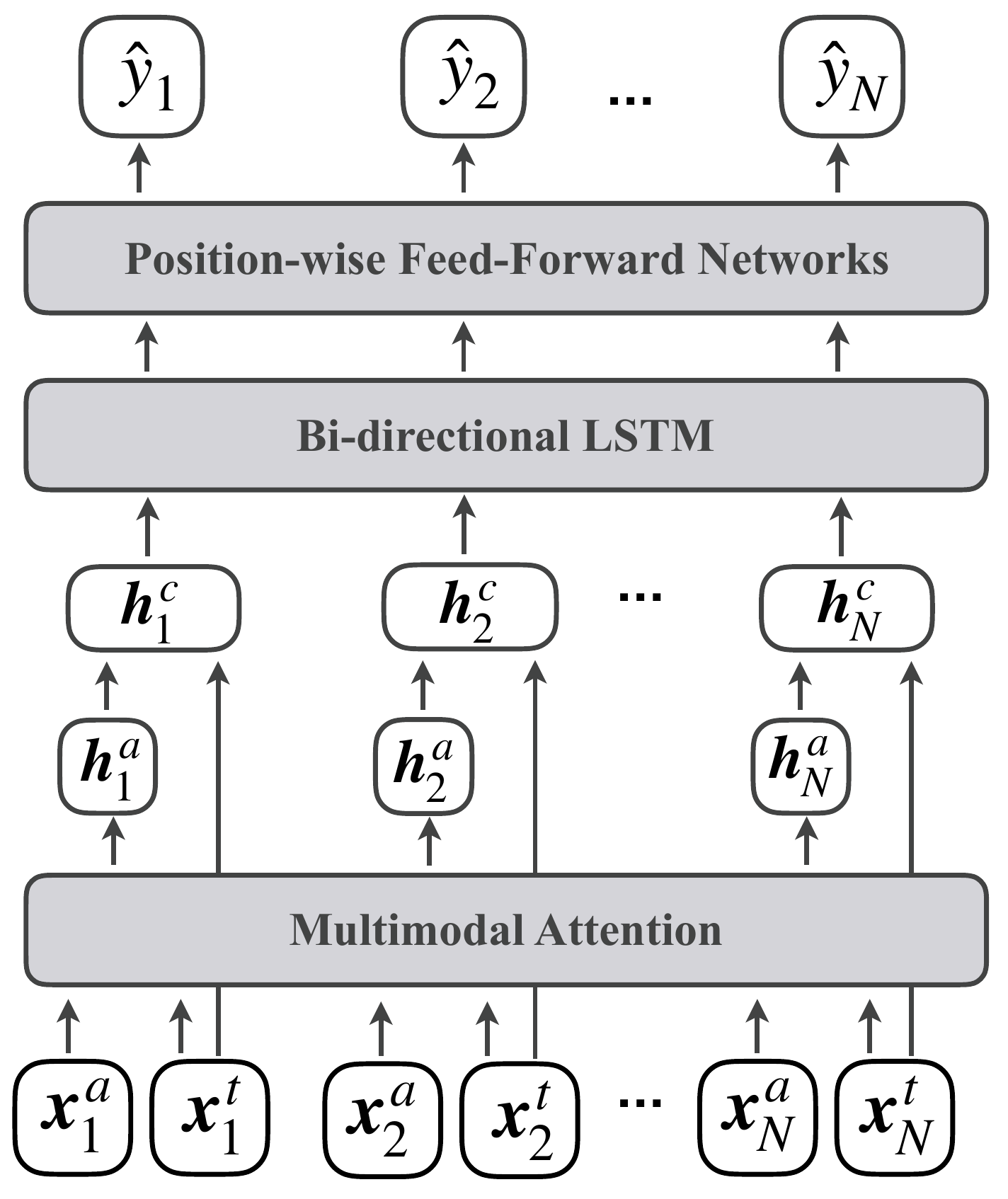}}
    \centerline{(b)}\medskip
\end{minipage}
    \caption{(a) Multimodal attention layer. (b) The proposed neural framework.}
    \label{fig:network_structure}
\end{figure}

In order to capture the information from both vocal and language modalities in the classroom environment, we design a novel multimodal attention layer that is able to alleviate the language ambiguity by using the voice attention mechanism. The majority of classroom conversations is open-ended and it is very difficult to distinguish its activity type when only considering the  sentence itself. Furthermore, not every piece of contextual segments contributes equally to the labeling task, especially the context is a mix of segments from teachers and students. Therefore, we use acoustic information as a complementing resource. More specifically, for each segment $s_i$, we not only utilize its own acoustic and language information but also the contextual information within the entire classroom recording. Moreover, the contextual segments are automatically weighted by the voice attention scores. The voice attention scores aim to cluster segments from the same subject (teacher or student), which is illustrated in Figure \ref{fig:network_structure}(a).

As shown in Figure \ref{fig:network_structure}(a), firstly, each segment is treated as a query and we compute its attention scores between the query and all the remaining segments by using the acoustic features. We choose the standard scaled dot-product as our attention function \cite{vaswani2017attention}. The scaled dot-product scores can be viewed as the substitutes of the cosine similarities between voice embedding vectors, which have been commonly used as a calibration for the acoustic similarity between different speakers' utterances \cite{wan2018generalized,wang2018speaker}. After that, we compute the multimodal representation by attending scores to contextual language features. The above process can be concisely expressed in the matrix form as $\boldsymbol{Q}=\boldsymbol{K}=\boldsymbol{X}^a$ and $\boldsymbol{V}=\boldsymbol{X}^t$ where $\boldsymbol{Q}$, $\boldsymbol{K}$ and $\boldsymbol{V}$ are the query, key and value matrices in the standard attention framework \cite{vaswani2017attention}. 

Both $\boldsymbol{X}^a$ and $\boldsymbol{X}^t$ are from pre-trained models. The acoustic feature $\boldsymbol{X}^a$ is obtained from the state-of-the-art LSTM-based speaking embedding network proposed by Wan et al. \cite{wan2018generalized}. The d-vector generated by such network has been proven to be effective in representing the voice characteristics of different speakers in speaker verification and speaker diarization problems \cite{wang2018speaker}. The language feature $\boldsymbol{X}^t$ comes from the word embedding network proposed by Mikolov et al. \cite{mikolov2013efficient}, which is widely used in many neural language understanding models \cite{vaswani2017attention,huang2015bidirectional}. In practice, to achieve better performance on the classroom specific datasets, we also fine tune the pre-trained models with linear transformation operators, i.e.,  $\boldsymbol{Q}=\boldsymbol{K}=\boldsymbol{X}^a \boldsymbol{W}^a; \quad \boldsymbol{V}=\boldsymbol{X}^t \boldsymbol{W}^t$, where $\boldsymbol{W}^a \in\mathbb{R}^{d_a \times d_q}$ and $\boldsymbol{W}_t \in\mathbb{R}^{d_t \times d_v}$ are the linear projection matrices. 

The attention score matrix $\boldsymbol{A}\in\mathbb{R}^{N\times N}$ is computed through dot-product of $\boldsymbol{Q}$ and $\boldsymbol{K}$, and the softmax function is then used to normalized the score values. Finally, the output embedding matrix $\boldsymbol{H}$ is calculated by the dot product of $\boldsymbol{A}$ and the value matrix $\boldsymbol{V}$. The complete equation is shown as follow:

\begin{equation}
\boldsymbol{H}^{a}=\mathrm{Attention}(\boldsymbol{Q},\boldsymbol{K},\boldsymbol{V})=\mathrm{softmax}(\cfrac{\boldsymbol{Q}\boldsymbol{K}^{T}}{\sqrt{d_{q}}})\boldsymbol{V} \nonumber
\end{equation}

With our multimodal attention layer, the acoustic features $\boldsymbol{X}^a$ served as a bridge that wisely connects the scattered semantic features $\boldsymbol{X}^t$ across different segments. 

\subsubsection{The Overall Architecture}

By utilizing the above multimodal attention layer, we are able to learn the fused multimodal embeddings for each segment. Similar to \cite{he2016deep}, we add a residual connection by concatenating $\boldsymbol{h}_i^a$ with $\boldsymbol{x}^t_i$ in multimodal attention block, i.e., $\boldsymbol{h}_i^c = [\boldsymbol{h}_i^a;\boldsymbol{x}^t_i]$ and $\boldsymbol{H}^{c} \in \mathbb{R}^{N \times 2d_v}$ is the matrix form of all $\boldsymbol{h}_i^c$s. What's more, to better capture the contextually sequential information within the entire classroom recording, we integrate the position information of the sequence by using a Bi-directional LSTM (BiLSTM) layer after the residual layer \cite{huang2015bidirectional}. We denote the hidden representation of BiLSTM as $\boldsymbol{H}^{b}$, where $\boldsymbol{H}^{b} = \mathrm{BiLSTM}(\boldsymbol{H}^{c})\in\mathbb{R}^{N\times 2d_b}$ and $2d_b$ is the size of hidden vector of $\mathrm{BiLSTM}$. Finally, we use a two-layer fully-connected position-wise feed forward network to conduct the final predictions, i.e., $\mathcal{\hat{Y}} = \mathrm{softmax}(\mathrm{FCN}(\boldsymbol{H}^{b}))$, where $\mathrm{softmax}(\cdot)$ denotes the softmax function and $\mathrm{FCN}(\cdot)$ denotes the fully-connected network. The entire framework is illustrated in Figure \ref{fig:network_structure} (b).

In our multimodal learning framework, we use binary cross-entropy loss $\mathcal{L}_{c}$ to optimize the prediction accuracy, which is defined as $\mathcal{L}_{c}=\sum_{i=1}^{N}\ y_{i}\log \hat{p}_{i} + (1-y_{i})\log (1-\hat{p}_{i})$, where $\hat{p}_{i}$ the prediction probability for segment $s_i$.

Besides that, we also want to optimize the multimodal attention scores with existing label information. Therefore, we introduce an attention score regularization operator $\mathcal{R}_{\alpha}$. $\mathcal{R}_{\alpha}$ aims to penalize the similarity scores between two segments when they are from different activity types. $\mathcal{R}_{\alpha}$ is defined as $\mathcal{R}_{\alpha} = \sum_{i=1}^{N}\sum_{j=1}^{N}\alpha_{ij}\cdot\mathbf{1}_{\{y_{i}\ne y_{j}\}}$, where $\alpha_{ij}$ is the ($i,j$)th element in $\boldsymbol{A}$ and represents the attention score between $s_i$ and $s_j$. $\mathbf{1}_{\{\cdot\}}$ is the indicator function.

Therefore, the final loss function $\mathcal{L}$ in our multimodal learning framework is shown as $\mathcal{L}=\mathcal{L}_{c}+\beta \mathcal{R}_{\alpha}$, where $\beta$ is the hyper parameter and is selected (in all experiments) by the internal cross validation approach while optimizing models' predictive performances.

%% file: statement.tex
Let $\mathcal{S}$ be the sequence of segments of a classroom recording, i.e., $\mathcal{S} = \{s_i\}_{i=1}^N$ where $s_i$ denotes the \emph{i}th segment and $N$ is the total number of segments. Let $\mathcal{Y}$ be the corresponding label sequence, i.e., $\mathcal{Y} = \{y_i\}_{i=1}^N$ and each $y_i$ represents the classroom activity type, i.e, whether the segment is spoken by a student or a teacher. For each segment $s_i$, we extract both the acoustic feature $\boldsymbol{x}^{a}_{i} \in \mathbb{R}^{d_a \times 1}$ and the text feature $\boldsymbol{x}^t_i \in \mathbb{R}^{d_t \times 1}$. $d_a$ and $d_t$ are the dimensionality of $\boldsymbol{x}^{a}_{i}$ and $\boldsymbol{x}^{t}_{i}$. Let $\mathbf{X}^a$ and $\mathbf{X}^t$ be the acoustic and text feature matrices of sequence $\mathcal{S}$, i.e., $\mathbf{X}^a \in \mathbb{R}^{N \times d_t}$ and $\mathbf{X}^t \in \mathbb{R}^{N \times d_t}$. With the aforementioned notations and definitions, we can now formally define the CAD problem as a sequence labeling problem:

\textit{Given a classroom recording segment sequence $\mathcal{S}$ and the corresponding acoustic and text feature matrices $\mathbf{X}^a$ and $\mathbf{X}^t$, the goal is to find the most probable classroom activity type sequence $\mathcal{Y}$ as follows: }

\begin{equation}
    \mathcal{\hat{Y}} = \mathop{\arg\max}_{\mathcal{Y}\in \mathcal{\boldsymbol{Y}}} P(\mathcal{Y}|\mathbf{X}^a, \mathbf{X}^t) \nonumber
\end{equation}

\noindent where $\mathcal{\boldsymbol{Y}}$ is the collection of all possible labeling sequences and $\mathcal{\hat{Y}}$ is the predicted classroom activity type sequence.

%% file: experiment.tex
\subsection{Experimental Setup}
\label{sec:setup}
\input{setup}

\subsection{Baselines}
\input{baselines}

\subsection{Datasets}
\input{datasets}

\subsection{Results \& Analysis}
\input{results}

%% file: setup.tex
In the following experiments, we first feed each classroom recording to a publicly available third-party ASR online transcription service to (1) generate the unlabeled segment sequence; (2) filter out silence or noise segments; and (3) obtain the raw sentences transcriptions. Then similar to \cite{wang2018speaker}, each segment level audio signal is transformed into frames of width 25ms and step 10ms, and log-mel-filterbank energies of dimension 40 are extracted from each frame. After that, we build sliding windows of a fixed length (240ms) and step (120ms) on these frames. We run the pre-trained acoustic neural network on each window and compute acoustic features, i.e., $\boldsymbol{x}^a_i$, by averaging these window level embeddings. Similarly, text features, i.e., $\boldsymbol{x}^t_i$, are also generated from a pre-trained word embedding network.

The projected dimension, i.e., $d_q$, and the number of neurons in BiLSTM, i.e., $d_b$ are set to 64 and 100. The numbers of neurons in the final two-layer fully connected network are 128 and 2. We use ReLU as the activation function. We set $\beta$ to 10 and use ADAM optimizer with learning rate of 0.001. We set batch size and the number of training epoch to 64 and 20, respectively. 

We compare with different methods by accuracy and F1 score. Different from standard classification evaluation in which examples are equal, in CAD tasks, the lengths of different segments vary a lot. Therefore, evaluation results are weighted by the time span of each segment. The weight is computed as the proportion of each segment's duration over the sum of total segment durations of the class recording.

%% file: baselines.tex
We carefully choose the following state-of-the-art CAD related approaches as our baselines. 
They are: 
(1) \textbf{BiLSTM with acoustic features} ($\mathrm{BiLSTM}^{a}$): We train the BiLSTM model with acoustic features only. The text features are completely ignored and there is no multimodal attention fusion; 
(2) \textbf{BiLSTM with text features} ($\mathrm{BiLSTM}^{t}$): Similar to $\mathrm{BiLSTM}^{a}$ but instead only the text features are used; 
(3) \textbf{attention based BiLSTM with acoustic features} (Attn-$\mathrm{BiLSTM}^{a}$): A self-attention layer is added before BiLSTM model, and similar to $\mathrm{BiLSTM}^{a}$, only acoustic features are used; 
(4) \textbf{attention based BiLSTM with text features} (Attn-$\mathrm{BiLSTM}^{t}$): Similar to Attn-$\mathrm{BiLSTM}^{a}$ but instead only the text features are used;
(5) \textbf{BiLSTM with concatenated features} ($\mathrm{BiLSTM}^{c}$): Both the acoustic and text features are used and the concatenation of them are fed into the BiLSTM model directly without multimodal attention fusion; 
(6) \textbf{spectral clustering with d-vectors} ($\mathrm{Spectral}$): Spectral clustering on speaker-discriminative embeddings (a.k.a. d-vectors) \cite{wang2018speaker}. It first extracts d-vectors from acoustic features, and then applies the spectral clustering to cluster all the segments. After that, a classifier takes both acoustic and text features and predict the final activity type; 
(7) \textbf{unbounded interleaved-state recurrent neural networks} ($\mathrm{UIS-RNN}$): Similar to \emph{Spectral}, but instead of using spectral clustering, UIS-RNN uses a  distance-dependent Chinese restaurant process \cite{zhang2019fully}.

%% file: datasets.tex
To assess the proposed framework, we conduct several experiments on two real-world K-12 education datasets. 

\noindent \textbf{Online Classroom Data}(``Online'') We collect 400 online class recordings from a third-party online education platform. The data is recorded by webcams via live streaming. After generating segment sequences according to steps in Section \ref{sec:setup}, we label each segment as either $Teacher$ or $Student$. For those segment consisting of both teacher speaking and student speaking, we label it as the dominant activity type in the segment. The average duration of classroom recordings is 60 minutes. The average length of the segment sequences is 700. We train our model with 350 recordings and use the remaining 50 recordings as the test set. 

\noindent \textbf{Offline Classroom Data}(``Offline"): We also collect another 50 recordings from offline classroom environment as an additional test set. The data is obtained by indoor cameras that are installed on the ceiling of the classrooms.

%% file: results.tex
The results show that our approach outperforms all other methods on both \emph{Online} and \emph{Offline} datasets. Specifically, from Table \ref{tab:results}, we find the following results: (1) when comparing acoustic feature only models ($\mathrm{BiLSTM}^{a}$, Attn-$\mathrm{BiLSTM}^{a}$) to text feature only models ($\mathrm{BiLSTM}^{t}$, Attn-$\mathrm{BiLSTM}^{t}$), we can see that models based on acoustic features in general perform worse than models based on text features. We believe this is because two similar voice segments may have different activity types but the corresponding spoken terms may differ; (2) comparing $\mathrm{BiLSTM}^{a}$ and Attn-$\mathrm{BiLSTM}^{a}$, $\mathrm{BiLSTM}^{t}$ and Attn-$\mathrm{BiLSTM}^{t}$, blindly incorporating attention mechanism cannot improve the performance in CAD tasks due to the fact that the sequence is mixed by teacher spoken segments and student spoken segments; (3) the performance on teacher is better than student in general, this is because the majority of segments in the classroom recording is spoken by teachers. The percentages of student talk time is 22\% on both \emph{Online} and \emph{Offline} datasets.

\input{table}

%% file: table.tex
\begin{table}[!hptb]
\caption{Experimental results on \emph{Online} and \emph{Offline} datasets. $\mathrm{F1}_{T}$ and $\mathrm{F1}_{S}$ indicate the F1 scores for activity types \emph{Teacher} and \emph{Student}.}
\centering
\begin{small}
\begin{tabular}{@{}lccccccc@{}} \toprule
\label{tab:results}
    & \multicolumn{3}{c}{\textbf{Online}} & \multicolumn{3}{c}{\textbf{Offline}} \\
    & $\mathrm{Acc}$ & $\mathrm{F1}_{T}$ & $\mathrm{F1}_{S}$ & $\mathrm{Acc}$ & $\mathrm{F1}_{T}$ & $\mathrm{F1}_{S}$ \\ \cline{2-4} \cline{5-7}
    $\mathrm{BiLSTM}^{a}$        & 0.76 & 0.84 & 0.54 & 0.72 & 0.82 & 0.35 \\
    Attn-$\mathrm{BiLSTM}^{a}$   & 0.73 & 0.80 & 0.58 & 0.77 & 0.86 & 0.28 \\
    $\mathrm{BiLSTM}^{t}$        & 0.83 & 0.90 & 0.48 &	0.78 & \bf{0.87} & 0.28 \\
    Attn-$\mathrm{BiLSTM}^{t}$   & 0.78 & 0.87 & 0.13 &	0.78 & \bf{0.87} & 0.03 \\
    $\mathrm{BiLSTM}^{c}$        & 0.81 & 0.88 & 0.61 & 0.79 & \bf{0.87} & 0.24 \\
    Spectral                     & 0.78 & 0.86 & 0.53 &	0.73 & 0.83 & 0.37 \\
    UIS-RNN                      & 0.81 & 0.88 & 0.58 &	0.71 & 0.81 & 0.35 \\
    Our                          & \bf{0.92} & \bf{0.95} & \bf{0.81} & \bf{0.80} & \bf{0.87} & \bf{0.48} \\ \bottomrule
\end{tabular}
\end{small}
\end{table}

%% file: conclusion.tex

In this paper, we presented an attention based multimodal learning framework for CAD problems. Our approach is able to fuse data from different modalities and capture the long-term semantic dependence without any additional recording device. Experiment results demonstrated that our approach outperforms other state-of-the-art CAD learning approaches in terms of accuracy and F1 score.